%% file: conference_101719.tex
\documentclass[conference]{IEEEtran}
\IEEEoverridecommandlockouts
\usepackage{cite}
\usepackage{amsmath,amssymb,amsfonts}
\usepackage{pifont}
\usepackage{algorithmic}
\usepackage{graphicx}
\usepackage{subcaption}   
\usepackage{textcomp}
\usepackage{xcolor}
\usepackage{hyperref}
\usepackage{booktabs}
\usepackage{array}
\usepackage{tikz}
\usetikzlibrary{shapes.geometric, arrows.meta, positioning, calc, fit, backgrounds}
\def\BibTeX{{\rm B\kern-.05em{\sc i\kern-.025em b}\kern-.08em
    T\kern-.1667em\lower.7ex\hbox{E}\kern-.125emX}}

\begin{document}

\title{
Catching Transpilation Drift with a CI/CD Workflow in Quantum Software Development\\
\thanks{This work has been supported by the Academy of Finland (project DEQSE 349945), Business Finland (EM4QS 155/31/2024), Finnish Ministry of Education and Culture through the Quantum Doctoral Education Pilot Program (QDOC VN /3137/2024-OKM-4) and the Research Council of Finland through Finnish Quantum Flagship project (359240, JYU).
}
}

\author{\IEEEauthorblockN{Jouni Peltonen}
\IEEEauthorblockA{University of Jyväskylä \\
Jyväskylä, Finland \\
jouni.s.peltonen@jyu.fi}
\and
\IEEEauthorblockN{Otso Kinanen}
\IEEEauthorblockA{University of Jyväskylä \\
Jyväskylä, Finland \\
otso.j.r.kinanen@jyu.fi}
\and
\IEEEauthorblockN{Vlad Stirbu}
\IEEEauthorblockA{University of Jyväskylä \\
Jyväskylä, Finland \\
vlad.a.stirbu@jyu.fi}
}

\IEEEoverridecommandlockouts

\IEEEpubid{\parbox{\columnwidth}{\footnotesize © 2026 IEEE. Personal use of this material is permitted. Permission from IEEE must be obtained for all other uses.\hfill}\hspace{\columnsep}\makebox[\columnwidth]{}}

\maketitle

\begin{abstract}
Quantum software workflows rely on compiler and provider toolchains that evolve independently of application source code. 
Consequently, an unchanged quantum circuit may transpile into a different target-specific realization after changes in SDK versions, optimization settings, basis gates, coupling maps, or backend descriptions. 
Such \emph{transpilation drift} can affect circuit depth, gate composition, qubit mapping, and execution behavior, yet it is rarely monitored in CI/CD pipelines.

This paper proposes a Quantum DevOps workflow for detecting transpilation drift before execution. 
The workflow transpiles source circuits against configured target profiles, computes structural drift metrics, records provenance and artifacts in MLflow, and raises configurable warnings or failures in GitHub Actions. 
Using representative circuits and target profiles, we show how drift checks can expose toolchain-induced changes and support reproducibility audits. 
The contribution is a practical CI/CD guardrail for making quantum compilation behavior observable, testable, and auditable.
\end{abstract}


\section{Introduction}

Continuous integration and delivery (CI/CD) pipelines are widely used to detect software regressions early and to automate the path from source changes toward deployment\cite{elbaum2014techniques}. 
In quantum software projects, however, these pipelines typically validate source-level code, unit tests, or simulator outputs, while leaving an important regression source largely unmonitored: changes in the transpiled circuit produced by evolving quantum toolchains\cite{automated2024saxena, fingerhuth2018open}.

Quantum programs are rarely executed in the form written by developers. 
Before execution, source circuits are transformed by compiler and provider-specific workflows that adapt them to a target backend. 
Even when the source circuit is unchanged, updates to SDK versions, backend target descriptions, coupling maps, optimization settings, or provider integration layers can change the realized circuit. 
We call this phenomenon \emph{transpilation drift}.

Transpilation drift is not necessarily a correctness bug or evidence of ideal semantic non-equivalence. 
Rather, it is a software-engineering signal: the circuit that would be executed has changed, and the change may affect reproducibility, comparability, execution cost, or observed behavior. 
Without automated drift detection, such changes may only become visible after failed experiments, unexpected measurement distributions, or inconsistent cross-provider results.

In quantum workflows, deployment ultimately means submitting circuits to quantum hardware, where execution is costly, queue-based, and calibration-dependent. 
Detecting transpilation drift before this step therefore serves as a release gate: it avoids committing scarce hardware resources to circuits whose realization has silently changed, and flags when a result may no longer be comparable with a previously deployed baseline.

Following an objective-centered design science approach~\cite{dsrm}, we propose a lightweight Quantum DevOps workflow that transpiles circuits to configured target profiles, compares the resulting artifacts against a stored baseline, logs metrics and artifacts to MLflow\footnote{https://github.com/mlflow/mlflow}, and raises configurable warnings or failures in GitHub Actions\footnote{https://docs.github.com/en/actions}. We demonstrate it across selected hardware targets, compilation strategies, and circuits, and evaluate the sources of drift in the demonstration cases. The primary check operates at the transpilation-artifact level because it is fast, repeatable, and suited to pull-request (CI) workflows; execution-level checks are supported as optional scheduled or release-gate (CD) checks before circuits are deployed to hardware.

For this paper we set three objectives:
\begin{itemize}
    \item Objective 1: \textbf{Detectability} - detect transpilation drift before hardware execution.
    \item Objective 2: \textbf{Auditability} - preserve provenance/artifacts for reproducibility.
    \item Objective 3: \textbf{Actionability} - turn drift into CI feedback: pass/warn/fail. 
\end{itemize}

\section{Background and Related Work}
\label{relatedwork}

\textbf{Quantum DevOps and CI/CD.}
Software engineering practices such as continuous integration, automated testing, and deployment are increasingly studied in the quantum context\cite{kinanen2025toolchain, Gheorghe-Pop_Tcholtchev_Ritter_Hauswirth_2020}.
Existing efforts emphasize pipeline automation and program-level testing; the compiled, target-specific artifact is generally treated as a transient build product rather than a tracked CI output.
Our work complements these efforts by adding the transpiled artifact itself as a monitored CI output.

\textbf{Transpilation and compilation variability.}
Quantum transpilation is target- and toolchain-dependent\cite{yan2024quantum}.
Benchmarking studies have shown that compiler choices and versions substantially affect circuit structure and resulting quality\cite{Quetschlich_2023, stefano2024empirical}.
These studies characterize compilation behavior offline; we instead operationalize structural comparison as an automated, baseline-relative check inside CI.

\textbf{Reproducibility and experiment tracking.} 
Reproducibility of quantum experiments depends on recording the precise software and execution context~\cite{fingerhuth2018open}. 
Experiment-tracking systems such as MLflow are widely used in classical machine learning to log parameters, metrics, and artifacts for later comparison. An adaptation of the experiment-tracking for quantum software development suggests that quantum-specific datasets should be collected during the development process \cite{kinanen2025experiment}.
We apply this practice to quantum compilation by recording transpilation provenance and drift metrics, enabling retrospective auditing of when and why a circuit realization changed.

\section{Transpilation Drift as a CI/CD Concern}
\label{sec:drift}

In classical CI/CD, regressions are usually associated with source-code changes, dependency updates, or configuration changes. 
Quantum software adds another regression surface: the compiled circuit representation itself. 

The transpilation process comprises several actions that affect the output: \textit{gate synthesis} (mapping to the target's supported gate set), \textit{gate mapping} (arranging qubits to match hardware connectivity), \textit{gate optimization} (reducing depth and gate count to limit noise), and \textit{resource allocation} (fitting the circuit to hardware constraints such as qubit count). Even minor changes in any of these may noticeably affect code quality~\cite{stefano2024empirical}.

A pull request may leave the source circuit unchanged while modifying dependencies, transpiler settings, target backend descriptions, or provider integration code that changes the circuit submitted for execution. CI monitoring is thus practical: artifacts are generated quickly, stored deterministically under fixed seeds and toolchain versions, and compared before any QPU job is submitted. A drift alert then acts as an early warning that an experiment result, benchmark comparison, or deployment may no longer be comparable with the approved baseline.

The check is intentionally pre-execution focused. 
A drift warning does not imply that the new circuit is incorrect or ideally non-equivalent to the previous one, but it indicates that the target-specific realization changed enough to require review. 
Execution-level checks, such as comparing measurement distributions, can be added as scheduled or release-gate jobs, but are not required for pull-request-level drift detection.

\section{Workflow Design and Prototype}

\begin{figure}[t]
\centering
\scriptsize
\begin{tikzpicture}[
    node distance=0.3cm and 0.7cm,
    box/.style={rectangle, rounded corners, draw=black!60, fill=white,
                align=center, minimum width=3.4cm, minimum height=0.7cm,
                inner sep=3pt},
    trig/.style={box, draw=blue!60, fill=blue!8},
    dec/.style={diamond, aspect=2.2, draw=black!60, fill=white,
                align=center, inner sep=1pt, minimum width=3.0cm},
    store/.style={cylinder, shape border rotate=90, aspect=0.25,
                  draw=violet!60, fill=violet!10, align=center,
                  minimum width=2.2cm, minimum height=1.5cm, inner sep=2pt},
    pass/.style={box, draw=green!55!black, fill=green!12, minimum width=1.4cm},
    warn/.style={box, draw=orange!80!black, fill=orange!12, minimum width=1.4cm},
    fail/.style={box, draw=red!70!black, fill=red!12, minimum width=1.4cm},
    flow/.style={-{Stealth[length=2mm]}, thick},
    log/.style={-{Latex[length=2mm, open]}, dashed, draw=violet!70}
]

\node[trig] (trigger) {Git event: PR / schedule / dispatch};
\node[box, below=0.7cm of trigger] (load) {Load drift config\\\textit{circuits, targets, thresholds}};
\node[box, below=of load] (discover) {Discover source circuits\\\textit{*.py / *.qasm}};
\node[dec, below=of discover] (foreach) {For each\\circuit $\times$ target};
\node[box, below=of foreach] (transpile) {Transpile (Qiskit)\\\textit{basis, coupling, opt.\ level, seed}};
\node[box, below=of transpile] (metrics) {Extract metrics\\\textit{depth, gates, 2Q, layout}};
\node[box, below=of metrics] (compare) {Compare vs.\ baseline\\compute $D_{\mathrm{struct}}$};
\node[dec, below=of compare] (threshold) {Threshold\\policy};

\node[store, right=1.6cm of compare] (mlflow) {MLflow\\baseline\\artifacts\\metrics};

\node[warn, below=1.0cm of threshold] (warn) {warn};
\node[pass, left=0.5cm of warn] (pass) {pass};
\node[fail, right=0.5cm of warn] (fail) {fail};

\node[trig, below=0.6cm of warn] (report) {Report to GitHub\\\textit{PR summary / build status}};

\begin{scope}[on background layer]
  \node[draw=black!45, dashed, rounded corners, fill=black!3,
        fit=(load)(discover)(foreach)(transpile)(metrics)(compare)(threshold),
        inner sep=10pt] (runner) {};
\end{scope}
\node[font=\scriptsize\bfseries, text=black!60, anchor=south west]
      at (runner.north west) {GitHub Actions runner};

\draw[flow] (trigger) -- (load);
\draw[flow] (load) -- (discover);
\draw[flow] (discover) -- (foreach);
\draw[flow] (foreach) -- (transpile);
\draw[flow] (transpile) -- (metrics);
\draw[flow] (metrics) -- (compare);
\draw[flow] (compare) -- (threshold);

\draw[flow] (threshold.south) -- (pass.north)
      node[pos=0.55, left=1pt, font=\tiny] {$<\!\tau_{\mathrm{w}}$};
\draw[flow] (threshold.south) -- (warn.north);
\draw[flow] (threshold.south) -- (fail.north)
      node[pos=0.55, right=1pt, font=\tiny] {$\geq\!\tau_{\mathrm{f}}$};

\draw[flow] (pass.south) |- (report.west);
\draw[flow] (warn.south) -- (report.north);
\draw[flow] (fail.south) |- (report.east);

\draw[log] (metrics.east) -| (mlflow.north)
      node[pos=0.3, above, font=\tiny, text=violet!70] {log run};
\draw[log] (mlflow.west) -- (compare.east)
      node[midway, above, font=\tiny, text=violet!70] {fetch baseline};
\draw[log] (threshold.east) -| (mlflow.south)
      node[pos=0.3, below, font=\tiny, text=violet!70] {log status};

\end{tikzpicture}
\caption{Transpilation-drift CI workflow. Solid arrows show the detection pipeline
(transpile $\rightarrow$ measure $\rightarrow$ compare-to-baseline $\rightarrow$
pass/warn/fail); dashed arrows show MLflow logging and baseline retrieval that
provide auditability. The threshold policy converts the structural drift score
$D_{\mathrm{struct}}$ into actionable CI feedback ($\tau_{\mathrm{w}}=\tau_{\mathrm{warn}}$,
$\tau_{\mathrm{f}}=\tau_{\mathrm{fail}}$).}
\label{fig:workflow}
\end{figure}
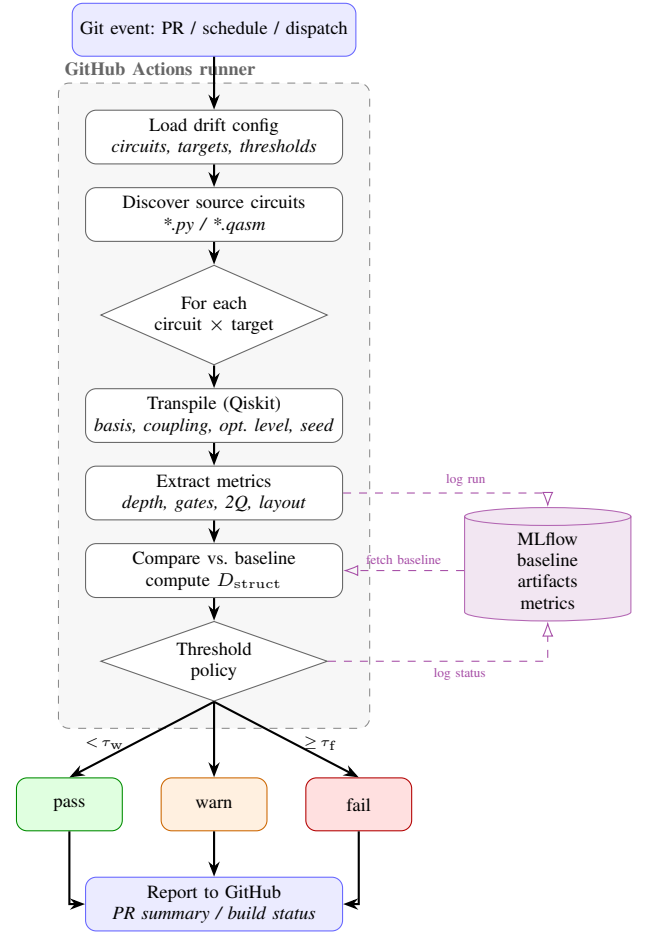

The proposed workflow is implemented as a GitHub Actions pipeline invoking a Python drift-analysis tool (Figure~\ref{fig:workflow}).
A repository-level configuration file specifies source circuits, target profiles, transpiler settings, baseline selection, and drift thresholds.

As shown in Figure~\ref{fig:workflow}, the pipeline (1) loads the circuits, targets, transpiler settings, and thresholds; (2) transpiles each circuit for each target profile; (3) measures structural metrics (depth, total gates, two-qubit gates, layout); (4) compares against a stored baseline; and (5) logs artifacts to MLflow and emits pass/warn/fail feedback in GitHub Actions.

Each CI run is logged as an MLflow run with parameters such as commit SHA, circuit identifier, target backend, basis gates, coupling map identifier, transpiler version, optimization level, and random seed. 
Artifacts include source circuits, transpiled circuits, backend target descriptions, metric JSON files, and plots.

For each source circuit \(c\), target profile \(t\), and CI run \(r\), the workflow records a structural feature vector of count-valued metrics:
\[
F(c,t,r)=\langle \mathrm{depth},\ \mathrm{gates},\ \mathrm{twoQ} \rangle .
\]
Qubit layout is recorded separately as provenance and flagged when it changes, rather than aggregated into the score.

Given a baseline run \(b\), the structural drift score is the mean relative change across these metrics:
\[
D_{\mathrm{struct}}(c,t,r,b)
=
\frac{1}{K}
\sum_{k=1}^{K}
\frac{|m_k(c,t,r)-m_k(c,t,b)|}
{\max\!\big(m_k(c,t,b),\,\epsilon\big)},
 \epsilon=1.
\]
The score is not a formal circuit distance; it is a CI signal indicating how much the compiled realization changed relative to the approved baseline. We use the mean for a stable aggregate signal; a max-based variant, which flags any single large metric change, is a straightforward configuration alternative.
A score of \(0\) means no change in the tracked metrics, while larger values indicate proportionally larger structural change.

The CI status follows configurable thresholds:
\[
\mathrm{status} =
\begin{cases}
\mathrm{pass}, & D_{\mathrm{struct}} < \tau_{\mathrm{warn}} \\
\mathrm{warn}, & \tau_{\mathrm{warn}} \leq D_{\mathrm{struct}} < \tau_{\mathrm{fail}} \\
\mathrm{fail}, & D_{\mathrm{struct}} \geq \tau_{\mathrm{fail}} .
\end{cases}
\]

In the demonstration we use illustrative values \(\tau_{\mathrm{warn}}=0.1\) and \(\tau_{\mathrm{fail}}=0.5\); systematic threshold calibration is left to future work (Section~\ref{discussion}).

\textbf{Baseline management.} 
The baseline is an approved CI run whose transpiled artifacts and structural metrics are stored in MLflow and referenced by run identifier. 
A baseline is established when a maintainer accepts a transpilation result as the reference realization for a given circuit and target profile. 
When a change is intentional and approved, the baseline is updated, so future drift is measured against the new reference. This makes drift detection relative and auditable rather than dependent on absolute thresholds alone.

\section{Demonstration}
\label{demonstration}

We demonstrate the workflow with two complementary drift scenarios that arise without any change to the source circuit.
In \emph{Scenario A (seed drift)}, an experiment is re-run with the source circuit, configuration, transpiler, and SDK version all held fixed, but the transpiler seed left unset. Because Qiskit transpilation is stochastic when \texttt{seed\_transpiler} is not specified, the realized circuit can differ between otherwise identical runs. This case is particularly instructive as nothing in the source diff or configuration would alert a reviewer, yet the executed circuit changes, precisely the situation that motivates automated, CI-level detection. We illustrate it with Bernstein--Vazirani circuit (BV12) across three target profiles (IonQ Forte, IQM Emerald, Rigetti Cepheus).

In \emph{Scenario B (transpiler drift)}, the baseline uses IQM's Qiskit transpiler (\texttt{iqm-client} v34.0.4, supported up to Qiskit v2.1.2), and the workflow detects the drift from switching to Qiskit's built-in transpiler on a newer release (v2.4.2). This is a realistic scenario: when a new Qiskit version ships, provider-updated transpilers typically follow later, so moving to the built-in transpiler may be necessary. This scenario can also arise when a team wants to harmonize on a single transpiler across all gate sets. We use a 10-qubit Quantum Fourier transform (QFT10) targeted to IQM's 54-qubit native gate set (\texttt{IQMFakeAphrodite}, matching IQM Emerald), as it exposes the difference more clearly than the smaller BV12 circuit.
\input{combined_comparison}
\input{qft10_iqm}
\begin{figure*}
    \centering
    \begin{subfigure}[t]{0.30\linewidth}
        \vspace{0pt}          
        \centering
        \includegraphics[width=\linewidth]{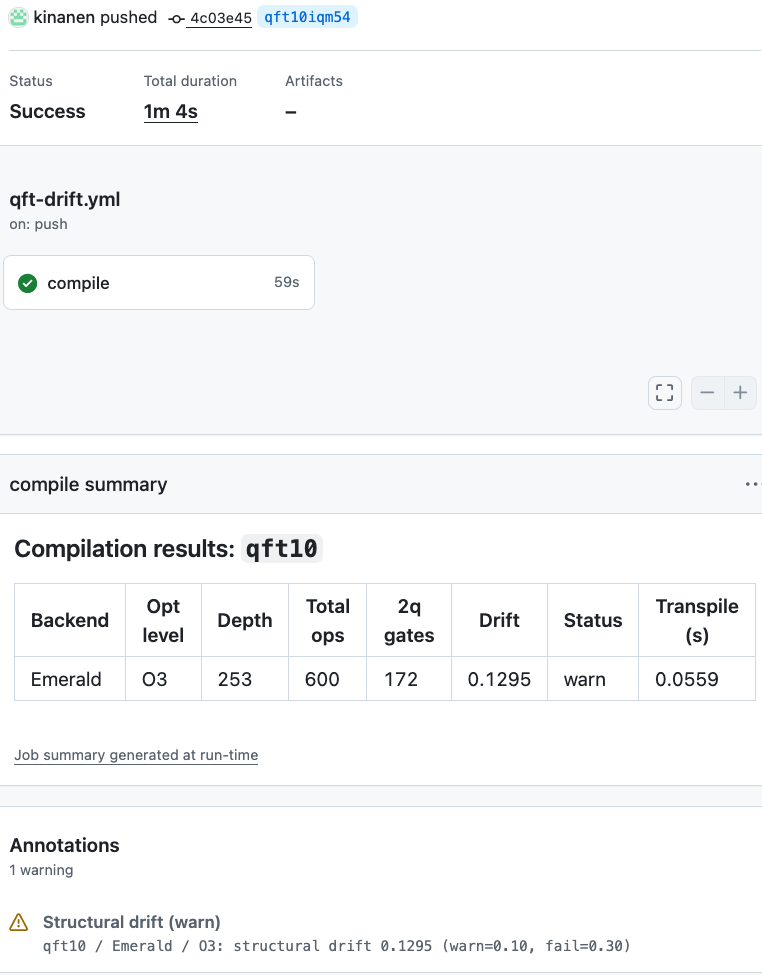}
        \caption{GitHub Actions workflow run.}
        \label{fig:mlflow-scenarioB-gh}
    \end{subfigure}
    \hfill                     
    \begin{subfigure}[t]{0.63\linewidth}
        \vspace{0pt}          
        \centering
        \includegraphics[width=\linewidth]{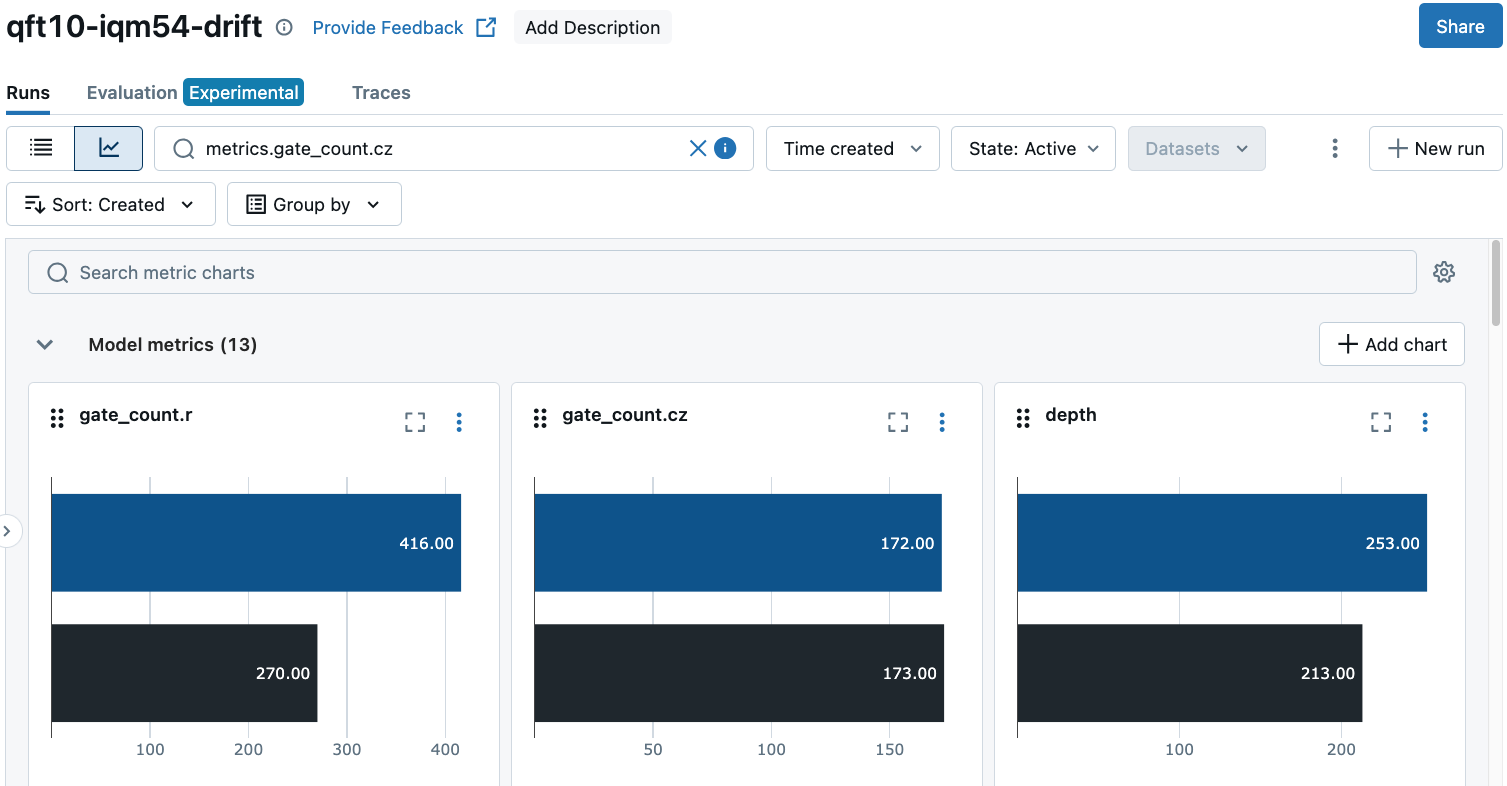}
        \caption{MLflow run comparison view.}
        \label{fig:mlflow-scenarioB-mlf}
    \end{subfigure}
    \caption{Scenario B (QFT10, IQM target): switching from IQM's \texttt{iqm-client} transpiler to Qiskit's built-in transpiler raises depth (213$\rightarrow$253) and yields a structural drift of 0.1295.(\subref{fig:mlflow-scenarioB-gh})~GitHub Actions CI views key metrics and shows transpilation drift warning. (\subref{fig:mlflow-scenarioB-mlf})~MLflow UI surfacing the per-run metrics, allowing to further examine the data;  Full per-gate metrics are given in Table~\ref{tab:qft10-transpiler-drift}.}
    \label{fig:mlflow-scenarioB}
\end{figure*}
Table~\ref{tab:demo-results} summarizes both scenarios and Table~\ref{tab:qft10-transpiler-drift} breaks down Scenario B, with corresponding GitHub and MLflow views in Figure~\ref{fig:mlflow-scenarioB}. In Scenario A, the seed affected only some targets in our runs: IonQ (all-to-all coupling) showed no drift, whereas IQM and Rigetti (limited connectivity) drifted, plausibly because constrained connectivity leaves more room for stochastic routing decisions. In Scenario B, the biggest change arises from the single qubit gate optimization, the transpilers also differ in qubit usage (IQM allocates 11 physical qubits, vanilla Qiskit 10); consistent with the workflow design (Section~\ref{sec:drift}), qubit layout is recorded as provenance and flagged separately rather than folded into the drift score.

The goal is not to benchmark providers, but to show that target-circuit changes can be detected, logged, and surfaced before hardware execution. Demo code and sample pipeline: \href{https://github.com/kinanen/quantum-transpilation-drift-detection}{github.com/kinanen/quantum-transpilation-drift-detection}.

\section{Discussion and Threats to Validity}
\label{discussion}

The proposed workflow treats transpilation drift as an engineering signal rather than a proof of incorrectness (Section~\ref{sec:drift}): structurally different circuits may implement the same ideal operation while differing in depth, gate composition, routing, cost, or noise exposure. Because the check is pre-execution by design, it suits pull-request workflows where hardware execution may be too costly, slow, or calibration-dependent; execution-level checks can be added as scheduled or release-gate jobs. Thresholds are project-specific and should be calibrated according to circuit family, target backend, and acceptable portability risk.

The current prototype demonstrates drift detection using a small circuit set and selected target profiles. 
Future work will expand target coverage, add longitudinal backend tracking, integrate execution-level checks, and evaluate threshold policies across larger quantum software repositories.

\textbf{Threats to validity.} 
The drift score is a structural proxy as it summarizes changes in depth and gate counts, and does not by itself establish semantic non-equivalence or predict execution fidelity. 
Because transpilation is performed client-side, measured drift reflects the combined behavior of the client toolchain and the target profile rather than provider-internal compilation alone. 
Threshold values are project-specific and require calibration. 
These limitations are acceptable for the intended purpose: providing an early, auditable signal that the compiled realization has changed, not a formal guarantee of correctness.

\section{Conclusion}
\label{conclusion}

This paper presented a CI/CD workflow for detecting transpilation drift in quantum software projects. 
The workflow supports three objectives: pre-execution detectability through baseline comparison of transpiled circuits, auditability through MLflow-based logging of artifacts and provenance, and actionability through configurable GitHub Actions warnings or failures. 
Rather than proposing a new compiler, hardware benchmark, or formal equivalence checker, the approach provides a lightweight Quantum DevOps guardrail for making compilation behavior observable and reviewable during development. 

\bibliographystyle{IEEEtran}
\bibliography{bibliography}
\end{document}

%% file: combined_comparison.tex
\begin{table}[t]
\centering
\caption{Example CI drift report across seed drift (A) and transpiler drift (B) scenarios. Deltas are relative to the approved baseline for each circuit/target.}
\label{tab:demo-results}
\footnotesize
\begin{tabular}{lllrrrl}
\toprule
Scenario & Circuit & Target & $\Delta$Depth & $\Delta$2Q & Drift & Status \\
\midrule
A: Seed       & BV12  & IonQ    & $+0$  & $+0$ & 0.000 & pass \\
A: Seed       & BV12  & IQM     & $+7$  & $+3$ & 0.160 & warn \\
A: Seed       & BV12  & Rigetti & $+5$  & $+3$ & 0.140 & warn \\
\midrule
B: Transpiler & QFT10 & IQM     & $+40$ & $-1$ & 0.1295 & warn \\
\bottomrule
\end{tabular}
\end{table}

%% file: qft10_iqm.tex
\begin{table}[t]
\centering
\caption{Quantum Fourier transform on 10 qubits, transpiled to IQM 54 qubit hardware. Vanilla Qiskit increases depth (+40) and single-qubit rotations (+148) while two-qubit count is essentially unchanged, illustrating that the IQM transpiler optimizes rotation gates more aggressively.}
\label{tab:qft10-transpiler-drift}
\footnotesize
\begin{tabular}{lrr}
\toprule
Metric & IQM (baseline) & Vanilla Qiskit \\
 & qiskit 2.1.2 & qiskit 2.4.2 \\
\midrule
Depth                  & 213 & 253 \\
CZ gates      & 173 & 172 \\
R gates       & 270 & 416 \\
RZ gates      & 0   & 2 \\
Qubits used           & 11  & 10 \\
Total ops             & 453 & 600 \\
2-qubit gates       & 173 & 172 \\
Structural drift      & -   & 0.1295 \\
\bottomrule
\end{tabular}
\end{table}

%% file: conference_101719.bbl
\begin{thebibliography}{10}
\providecommand{\url}[1]{#1}
\csname url@samestyle\endcsname
\providecommand{\newblock}{\relax}
\providecommand{\bibinfo}[2]{#2}
\providecommand{\BIBentrySTDinterwordspacing}{\spaceskip=0pt\relax}
\providecommand{\BIBentryALTinterwordstretchfactor}{4}
\providecommand{\BIBentryALTinterwordspacing}{\spaceskip=\fontdimen2\font plus
\BIBentryALTinterwordstretchfactor\fontdimen3\font minus \fontdimen4\font\relax}
\providecommand{\BIBforeignlanguage}[2]{{%
\expandafter\ifx\csname l@#1\endcsname\relax
\typeout{** WARNING: IEEEtran.bst: No hyphenation pattern has been}%
\typeout{** loaded for the language `#1'. Using the pattern for}%
\typeout{** the default language instead.}%
\else
\language=\csname l@#1\endcsname
\fi
#2}}
\providecommand{\BIBdecl}{\relax}
\BIBdecl

\bibitem{elbaum2014techniques}
S.~Elbaum, G.~Rothermel, and J.~Penix, ``Techniques for improving regression testing in continuous integration development environments,'' in \emph{Proceedings of the 22nd ACM SIGSOFT International Symposium on Foundations of Software Engineering}, 2014, pp. 235--245.

\bibitem{automated2024saxena}
M.~C. Saxena, A.~Tamrakar, and U.~Arranz, ``Automated testing and deployment strategies for quantum algorithms,'' in \emph{2024 7th International Conference on Contemporary Computing and Informatics (IC3I)}, vol.~7, 2024, pp. 771--779.

\bibitem{fingerhuth2018open}
M.~Fingerhuth, T.~Babej, and P.~Wittek, ``Open source software in quantum computing,'' \emph{PloS one}, vol.~13, no.~12, p. e0208561, 2018.

\bibitem{dsrm}
\BIBentryALTinterwordspacing
K.~Peffers, T.~Tuunanen, M.~A. Rothenberger, and S.~Chatterjee, ``A design science research methodology for information systems research,'' \emph{Journal of Management Information Systems}, vol.~24, no.~3, pp. 45--77, 2007. [Online]. Available: \url{https://doi.org/10.2753/MIS0742-1222240302}
\BIBentrySTDinterwordspacing

\bibitem{kinanen2025toolchain}
O.~Kinanen, A.~D. Mu{\~n}oz-Moller, V.~Stirbu, J.~M. Murillo, and T.~Mikkonen, ``Toolchain for faster iterations in quantum software development,'' \emph{Computing}, vol. 107, no.~4, pp. 1--28, 2025.

\bibitem{Gheorghe-Pop_Tcholtchev_Ritter_Hauswirth_2020}
I.-D. Gheorghe-Pop, N.~Tcholtchev, T.~Ritter, and M.~Hauswirth, ``Quantum devops: Towards reliable and applicable nisq quantum computing,'' in \emph{2020 IEEE Globecom Workshops (GC Wkshps}, 2020, pp. 1--6.

\bibitem{yan2024quantum}
G.~Yan, W.~Wu, Y.~Chen, K.~Pan, X.~Lu, Z.~Zhou, Y.~Wang, R.~Wang, and J.~Yan, ``Quantum circuit synthesis and compilation optimization: Overview and prospects,'' \emph{arXiv preprint arXiv:2407.00736}, 2024.

\bibitem{Quetschlich_2023}
N.~Quetschlich, L.~Burgholzer, and R.~Wille, ``Mqt bench: Benchmarking software and design automation tools for quantum computing,'' \emph{Quantum}, vol.~7, p. 1062, July 2023.

\bibitem{stefano2024empirical}
M.~D. Stefano, D.~D. Nucci, F.~Palomba, and A.~D. Lucia, ``An empirical study into the effects of transpilation on quantum circuit smells,'' \emph{Empirical Software Engineering}, vol.~29, no.~3, p.~61, 2024.

\bibitem{kinanen2025experiment}
O.~Kinanen, M.~Gamage, and V.~Stirbu, ``Toolchain for experiment tracking in iterative quantum software development,'' in \emph{2025 IEEE International Conference on Quantum Computing and Engineering (QCE)}, vol.~2.\hskip 1em plus 0.5em minus 0.4em\relax IEEE, 2025, pp. 187--192.

\end{thebibliography}
